\RequirePackage{ifpdf}
\ifpdf 
\documentclass[pdftex]{sigma}
\else
\documentclass{sigma}
\fi

\begin{document}

\allowdisplaybreaks

\renewcommand{\thefootnote}{$\star$}

\renewcommand{\PaperNumber}{009}

\FirstPageHeading

\ShortArticleName{Introduction to Sporadic Groups}

\ArticleName{Introduction to Sporadic Groups\footnote{This
paper is a contribution to the Proceedings of the Workshop ``Supersymmetric Quantum Mechanics and Spectral Design'' (July 18--30, 2010, Benasque, Spain). The full collection
is available at
\href{http://www.emis.de/journals/SIGMA/SUSYQM2010.html}{http://www.emis.de/journals/SIGMA/SUSYQM2010.html}}}

\Author{Luis J.~BOYA}

\AuthorNameForHeading{L.J.~Boya}

\Address{Departamento de F\'{\i}sica Te\'orica, Universidad
de Zaragoza, 50009 Zaragoza, Spain}
\Email{\href{mailto:luisjo@unizar.es}{luisjo@unizar.es}}

\ArticleDates{Received September 18, 2010, in f\/inal form January 12, 2011;  Published online January 16, 2011}

\Abstract{This is an introduction to f\/inite simple groups, in
particular sporadic     groups, intended for physicists. After a
short review of group theory, we   enumerate the $1+1+16=18$
families of f\/inite simple groups, as an introduction    to the
sporadic groups. These are described next, in three levels of
increasing complexity, plus the six isolated ``pariah'' groups. The
(old) f\/ive Mathieu groups    make up the f\/irst, smallest order
level. The seven groups related to the Leech  lattice,    including
the three Conway groups, constitute the second level. The third and
highest level contains the Monster group $\mathbb M$, plus seven
other related groups. Next a brief mention is made of the remaining
six {\it pariah} groups, thus completing the $5+7+8+6=26$ sporadic
groups. The review ends up with     a brief     discussion of a few
of physical applications of f\/inite groups in physics,   including a
couple of recent examples which use sporadic groups.}


\Classification{20D08; 20F99}

\renewcommand{\thefootnote}{\arabic{footnote}}
\setcounter{footnote}{0}

\section{Introduction}

\subsection{Motivation} Finite groups were f\/irst applied in physics
to classify crystals (Bravais); with the advent of quantum mechanics
(1925), emphasis shifted towards continuous (Lie) groups  (Wigner,
Weyl). Around 1960 some groups, like ${\rm SU}(3)$ (f\/lavour) were employed
to classify particle states (Gell-Mann). Today one needs no
justif\/ication to use routinely Lie groups and their representations
in physics.

On the other hand, the use of discrete, f\/inite groups in particle
physics has been limited to the symmetric group ${\rm S}_n$ (statistics
for identical particles), and to some involutory operations like
CPT. Recently, however, the use of f\/inite groups in particle physics
has been rekindled, due to new developments; for example, the
Monster group $\mathbb M$, which is the largest sporadic group,
f\/irst def\/ined (ca.~1980) as automorphism group of a certain algebra,
was afterwards also built up (around 1984) with the use of vertex
operators, a construct typical of the physical super string theory.
A Japanese group (Eguchi et al., June~2010) has found relations of
the K3 (complex) surface, much used in compactif\/ication of extra
dimensions, with the Mathieu group ${\rm M}_{24}$, the last of the f\/ive
groups constituting the f\/irst level of the sporadic groups.

So the aim of this review is to introduce f\/inite simple groups, in
particular sporadic groups, to a physics audience. We anticipate
that the use of these f\/inite groups in physics is only expected to
increase, so at this time an introduction to the subject seems
justif\/ied. In this section we set the stage for the situation, by
recalling def\/initions and elementary properties of our objects, the
groups whose order is f\/inite. A nice historical review of groups in
physics can be found in~\cite{Bon}.

\subsection{Elementary def\/initions and properties}

We recall a group $G$ is a set with an internal operation $G \times
G\rightarrow G$ ($\{g, k\} \rightarrow  gk$) associative with unity~$I$ (or~Id) and inverse $g^{-1}$. $|G| = n$ denotes the number of
elements, the order of the group. If $gk=kg$, the group is called
Abelian. Among the inf\/inite ($n=\infty$) groups, Lie groups enjoy a~special position, as they can be expressed by a f\/inite number of
parameters (real numbers), the dimension of the underlying group
manifold: e.g.\ the rotation group ${\rm SO}(3)$ is a~continuous Lie group
with three (bounded) parameters~\cite{Wey}.

A morphism (homomorphism) $\mu : G \rightarrow  Q$ is the natural
map, obeying $\mu(gk) =  \mu(g) \mu  (k)$, for any $g, k \in  G$. A
subgroup $H\subset G$ is a subset which is group by itself.
Conjugation of~$g$ by~$h$, i.e., $h: g \rightarrow h\cdot g\cdot
h^{-1}$ is morphism; it splits $G$ into disjoint classes of conjugate
elements. A subgroup $H\subset G$ invariant under conjugation,
$gHg^{-1}=H$, is called a {\it normal} subgroup (invariant,
distinguished); if $A$ is Abelian, any subgroup is normal. A group
$G$ is {\it simple}, if the only normal subgroups are $G$ itself and
the identity $I$. We are interested in f\/inite simple groups:
\begin{gather*}
 G \ {\rm f\/inite\; simple}\Longleftrightarrow |G|<\infty, \quad {\rm and}\quad  H\subset G \ {\rm
 normal} \ \Longrightarrow \  H=G\quad \mbox{or}\quad H={\rm Id}.
\end{gather*}

The {\it period} of $g \in  G$ is the minimum $r$ with $g^r=I$;
period 2 elements are called {\it involutions}, $a = a^{-1}$. If $g$
is of period $m$, it and its powers generate the {\it cyclic group}
$Z_m$, Abelian and of order~$m$. All even-order groups have
involutions {\it a} (Cauchy) (pair $g\ne a$ with $g^{-1}: {\rm Id}$ and
involutions {\it a} form an even-order set). With a pair $H\subset
G$, if we form $gH$ for $G\ni g\notin H$, we have the {\it left
coset} due to~$g$; two such, $gH$ and $kH$, verify $|gH| = |kH| =
|H|$, so $G$ can be put as union of same-order cosets; similarly for
{\it right} cosets. Hence, the order of $H$ {\it divides} that of~$G$: Lagrange theorem; for a prime~$p$, the group~$Z_p$ is therefore
{\it simple}: it is the {\bfseries \itshape first family} of
{\bfseries \itshape finite simple groups}, and the only Abelian one. The
number of cosets is called the index of $H$ in $G$, written $[G:H]$.
$ gH \equiv Hg$ if $H\subset G$ is normal; one can then also {\it
multiply} cosets, forming the so-called {\it quotient group}, noted
$Q:= G/H$, of order the index, $|Q| = [G:H]$. Viceversa, in any
morphism $\mu: G\rightarrow Q$, the image $\mu(G)$ is a subgroup of
$Q$, and ${\rm Ker}\, \mu:=\mu^{ -1}({\rm Id}_Q)$ is a~normal subgroup of $G$. We
write $Q:=G/H$ as a short exact sequence
\begin{gather*}
  1\longrightarrow H \longrightarrow G \longrightarrow Q
  \longrightarrow 1 ,
\end{gather*}
meaning $H$ injects in $G$ (monomorphism), $G$ covers $Q$
(epimorphism) and exactness means e.g.\ the kernel $(H)$ in the map $
G\rightarrow Q$ is the image of previous map, $H\rightarrow   G$.

If $z$ in $G$ verif\/ies $g\cdot z\cdot g^{-1} = z$, i.e.\ is f\/ixed by
conjugation, is called central; $I={\rm Id}$ is {\it central}; centrals
are class by themselves. The set of central elements makes up a
normal subgroup, called the {\bfseries \itshape centre} (center) of the
group, noted $C_G$. ${\rm SU}(2)$, used as quantum mechanical rotation
group, has center $\pm I$, i.e., isomorphic to~$Z_2$.

If    $\mu: G \rightarrow  G'$ is one-to-one {\it onto}, it has an
inverse ($\exists\,\mu^{ -1}$), and is called {\it isomorphism};
groups related by isomorphism are not considered dif\/ferent, $G
\approxeq G'$. {\it Endomorphism} is a $\mu : G \rightarrow  G$; the
set of endomorphisms of an {\it Abelian} group $A$ forms a ring,
${\rm End}(A)$. An isomorphism $i: G \rightarrow  G$ is called {\it
automorphism}, and their set ${\rm Aut}(G)$ is a well def\/ined group for
any group $G$. Conjugation is an automorphism, called {\it inner}.
They form a normal subgroup, ${\rm Inn}(G)$ normal in ${\rm Aut}(G)$, and the
quotient ${\rm Aut}(G)/{\rm Inn}(G) := {\rm  Out}(G)$ is called the group of classes of
automorphisms. The natural map $G \rightarrow  G$ given by
conjugation is morphism, whose kernel is clearly the centre~$C_G$ of
the group. We summarize these def\/initions in the following diagram
$(G', {\rm Ab}(G)$ explained next):
\begin{gather*}
  \begin{array}{@{}ccccccc}
           \, & \, & C_G & \, & \, & \, & \, \\
           \, &\, & \downarrow & \, & \, & \, & \, \\
           G' & \rightarrow & G & \rightarrow & {\rm Ab}(G) & \, & \, \\
           \, & \, & \downarrow & \, & \, & \, & \, \\
           \, & \, & {\rm Inn}(G) & \rightarrow & {\rm Aut}(G) & \rightarrow &
           {\rm Out}(G)
         \end{array}   \qquad\qquad {\bf Diagram \ I}
\end{gather*}

The {\it commutator} of two elements $g$, $k$ is $g\cdot k\cdot
g^{-1}\cdot k^{-1}$, $= {\rm Id}$ if\/f $g$ and $k$ commute, $gk=kg$; all
commutators $\{g\cdot k\cdot g^{-1}\cdot k^{-1}\}$ generate a normal
subgroup of $G$, called the {\it commutator} or derived group, $G'$;
the quotient $G/G'$ is Abelian (as the kernel has ``all''
noncommutativity), and it is the maximal Abelian image of $G$,
called ${\rm Ab}(G)$. If $G=G'$ the group is called ``perfect''.

$H$ normal in $G$ is {\it maximal}, if there is no $K$ normal in $G$
with $H$ normal in $K$; the quotient $G/H$ is then simple, say
$Q_1$. Now if $S$ is normal maximal in $H$, $Q_2:=H/S$ is simple
again, etc. For any f\/inite group G the chain of simple quotient $\{
Q_i\}$ ends in ${\rm Id}$, when the last maximal normal subgroup is
already simple. The {\it Jordan--H\"older theorem} asserts that the
chain $\{Q_i\}$ is unique up to the order: it does not depend on the
(in general, non-unique) maximal normal subgroup chosen each time.
If the family $\{Q_i\}$ is Abelian (hence of the form $Z_p$), the
group $G$ is called {\it solvable}.  For the opposite case, if $G$
is simple the J-H chain is $\{G, {\rm Id}\}$.

For $K$, $G$ groups, in the set of the Cartesian product $K \times G$
one establishes a group composition law by $(k, g)\cdot(k', g') =
(k\cdot k', g\cdot g')$, called the {\it direct product}. For two
groups~$K$,~$Q$ and a morphism $\mu : Q \rightarrow  {\rm Aut}(K)$, one
def\/ines the $\mu$-{\it semidirect product}, written $K\odot Q$, or
$K \otimes_{\mu} Q$ by setting, in the Cartesian product set, the
group law related to the morphism: $(k, g) \cdot (k', g') = (kk'',
gg')$, where the {\it aut} $\mu (g)$ leads $k'$ to $\mu(g)k':=k''$.

For a set of $n$ symbols: $1, 2, 3, \dots, n$, a {\it permutation}
is a new order $(123\dots n) \rightarrow  (1'2'3'\dots n')$; their
totality makes up the symmetric group ${\rm Sym}_n = {\rm S}_n$, with $n!$
elements; it is said of {\it degree}~$n$. A {\it transposition} (12)
is a 2-cycle, e.g.\ $123\dots n \rightarrow  213\dots n$; any element
in ${\rm S}_n$ can be written as a product of commutative non-overlapping
cycles, e.g.\ $(12345678)\rightarrow (21453687)$ is written as
$(12)(345)(6)(78)$, and this cycle structure is invariant under
inner autos (conjugation), so it separates ${\rm S}_n$ into its classes, as
many as partitions of number $n$; for example, ${\rm S}_3$ has three
classes, $({\rm Id})=(1)(2)(3)$, $(12)(3)$ and $(123)$; ${\rm S}_5$ has ${\rm Part}(5)
= 7$ classes, etc. Any permutation is reached from ${\rm Id}$ by certain
number of transpositions, whose parity (number mod~2) is conjugation
invariant: hence, the even permutations, $n!/2$ in number, make up
the index-two subgroup ${\rm Alt}_n$, necessarily normal (a single coset)
with quotient ${\rm S}_n/{\rm Alt}_n = Z_2$. The symmetric group is the most
important of the f\/inite groups; an easy {\it theorem of Cayley}
assures that any group of order~$n$ can be seen as subgroup of a
certain ${\rm S}_n$.

An {\it extension} $G$ of $K$ (kernel) by $Q$ (quotient) is the
triple (exact sequence) $e$: $1 \rightarrow K \rightarrow  G
\rightarrow  Q \rightarrow 1$ or simply $G/K\equiv  Q$. All f\/inite
groups are so constructed, starting with~$K$ and~$Q$ simple.
Extensions $E$ of $G$ by a cyclic group $Z_n$, $E/G \equiv Z_n$, are
marked at times as~$G\cdot n$. Viceversa, {\it coverings} (or Schur
{\it coverings}) are extensions $F$ of a certain $Z_m$ by $G$, $F/Z_m
\equiv G$ and are noted~$m\cdot G$. This is notation of the Atlas~\cite{Atl}, which is becoming wide-spread. Given $(K, Q)$, the possible
extensions  $e:  E/K \equiv Q$ are classif\/ied f\/irst by the maps
$Q\rightarrow {\rm Out}(K)$ (whereas the semidirect product, see above,
requires $Q\rightarrow {\rm Aut}(K))$. Later we shall use the term {\it
ampliations}~$E$ from~$G$ in a looser sense, meaning only a
particular embedding of G as subgroup (not normal!) in a~bigger
group~$E$, so $G\subset E$.

An important chapter of the theory of groups, indispensable in
physics, is the study of {\bfseries \itshape representations} (Frobenius,
Schur); we shall not need much of these here, so a few def\/initions
and results will suf\/f\/ice. A (linear) representation of a group $G$
is a morphism $D: G\rightarrow {\rm GL}(V)$ of~$G$ on the group of
invertible matrices, as automorphism group of a vector space $V$
over a~f\/ield~$K$. Usually $K=\mathbb R$ or $\mathbb C$, the real or the complex
numbers; $D(G) = {\rm Id}$ def\/ines the {\it identical} representation. A~subspace $U \subset V$ def\/ines a subrepresentation if is $G$-stable,
so $D(G)U\subseteq U$. $D$ is {\bfseries \itshape irreducible} if only the
whole~$V$ def\/ines (sub)representation. $D$ on $V$ and $D'$ on $V'$
(of same~$G$) are equivalent if there is a map $A : V\rightarrow V'$
permuting $G: AD(g)\cdot x = D'(g)A\cdot x$. For f\/inite (or compact)
$G$, one can chose $D(G)\subseteq U(V)$, in the unitary group $U =
U(V)$. The search of inequivalent irreducible (unitary)
representations ({\it irreps}) of a given group $G$ is a formidable
industry, with plenty of applications in physics and mathematics.
For a f\/inite group, the number of {\it irreps} coincides with the
number~$r$ of classes of conjugate elements. The sum of squares of
the dimensions $d_i$ of the $r$ {\it irreps} is (Burnside) the order
of the group: $|G| =\sum_{i=1}^r  d_i^2$. For example, for $G=A$
Abelian all {\it irreps} are 1-dim., and there are $|A|$ of them.
The number of 1-dim {\it irreps} is $|G/G'= {\rm Ab}(G)|$, so simple
groups have only one, ${\rm Id}$. For example, for the ${\rm S}_4$ group, the
Burnside relation is $|{\rm S}_4|= 4! = 24 = 2\cdot 1^2 + 2\cdot 3^2 +
1\cdot 2^2$. Finally, $|G|:d_i$, that is, the dim's of the {\it
irreps} are divisors of $|G|$.

The {\it character} $\chi$   of a representation $D$ is the Trace of
the representative matrices, so  $\chi_D(g) = {\rm Tr}\, D(g)$. Equivalent
{\it irreps} have same character, as so have elements in same class:
$\chi(g) = \chi(hgh^{-1})$. One can characterize \cite{Thom} any
{\it irrep} (and even the very same group $G$) by its $r \times  r$
{\bfseries \itshape table of characters}.

This material is completely standard. We quote \cite{Ram,Hal} and \cite{Wig} as good textbooks for physicists. Modern
texts for pure mathematicians are \cite{Rob} and \cite{Bogo}.

\subsection{Actions of groups}

For a group $G$ {\it to act} in a space $X$, written $G
\circ\rightarrow  X$, we mean $G$ acts permuting the points $x,
y,\dots$ in $X$; so we write $g(x)=y$, and impose ${\rm Id}(x)=x$ and
$(gk)(x) = g(k(x))$. This is equivalent to a~morphism $\mu:
G\rightarrow {\rm Sym}(X)$, with ${\rm Sym}(X) = {\rm S}_n$ if $|X| = n$. We consider
only both $G$ and $X$ f\/inite. Action of groups on spaces as
transformations is the very {\it raison d'\^etre} of the groups (F.~Klein). A representation is a case of a group acting in a vector
space.

The action is called {\it effective}, if ${\rm ker} \,\mu$  is ${\rm Id}$. So
$K:= {\rm ker}\,\mu$ is called the {\it ineffectivity kernel}, and $G_1:=
G/{\rm ker}\,\mu$ acts {\it effectively} or {\it faithfully}. The image of
a point $x$ by all $G$, $G(x)$, is called the {\it orbit} of $x$
(under~$G$). {\it Belonging to an orbit} is an equivalence
relation, so~$X$ is a union of disjoint orbits. If there is only an
orbit, the action is called {\it transitive}, meaning any point in~$X$ can go to any other in~$X$ by the action of~$G$. The {\it
stabilizer} (little group in physics) of a point~$x$,~$G_x$, is the
subgroup leaving it f\/ixed, $g(x)=x$ for $g \in  G_x$; points in the
same orbit have conjugate stabilizers. Fixed points are orbits by
themselves. One proves easily that $|G| = |G(x)|\cdot |G_x|$ for any
point $x$; if $G$ is transitive in $X$ with ${\rm Id}$ as stabiliser, $|G|
= |X|$; the action $G \circ \rightarrow  X$ is then called {\it
regular}. Some examples follow:

1) Let ${\rm GL}_n({\mathbb R})$ be the group of invertible real $n \times
n$ matrices acting on the real vector space $V = {\mathbb R}^n$;
there are two orbits, the origin 0 and the rest: the origin is a f\/ixed
point, and the stabilizer of any other point (nonzero vector) is the
af\/f\/ine group in dimension $(n-1)$.

2) Let $G$ be any group and $X=G$ the same set, acted upon itself by
conjugation. The action is inef\/fective, with the center as
inef\/fectivity kernel; the orbits are the classes of conjugate
elements, and the center is also the set of f\/ixed points.

 3) Let $P$ be a regular pentagon and $D_5$ the dihedral group,
 of rotations (by $72^\circ$) and ref\/lections leaving $P$ invariant.
 The action is ef\/fective and transitive, with little group $Z_2$:
 ref\/lections through the selected point.

4) Let $\Omega_ m^+$ be the positive mass hyperboloid of elementary
particles, namely the set of timelike future vectors $p$ $(= p_\mu )$
in momentum space for a f\/ixed mass $m > 0$. Let $L$ be the
(homogeneous) connected Lorentz group, acting naturally in these 4-dim
vectors. The action is ef\/fective and transitive, with little group
isomorphic to the 3-dim rotation group ${\rm SO}(3)$ (or ${\rm SU}(2)$ if one
considers the (double, universal) covering group ${\rm SL}_2({\mathbb C}) = L^\sim$
of the Lorentz group $L = {\rm SO}_0(3, 1))$.

5) Consider the orthogonal group ${\rm O}(3)$ and its connected subgroup
${\rm SO}(3)$ acting by isometries (rotations) in the ordinary sphere
$S^2$. The action is ef\/fective and transitive, with stabilizer
isomorph to ${\rm O}(2)$ and ${\rm SO}(2)$ respectively. If the north pole is
selected as f\/ixed, ${\rm SO}(2)$ spans the parallels, with the south pole
also f\/ixed.

6) The isometry rotation group of the {\it icosahedron} ($=Y_3$) is
the alternative ${\rm Alt}_5$ group, with 60 elements; it is transitive in
the 12 vertices; stabilizer is then $Z_5$.

7) Let $G$ acts on itself by left translations, $g: k \rightarrow
g\cdot k$. The action is transitive, with $I$ as stabilizer (=~regular action).

\cite{Cox-1} is a good reference for groups in geometry, see also~\cite{Con-1}.

\subsection{Examples of f\/inite groups}

We introduced already ${\rm S}_n$, ${\rm Alt}_n$ and $Z_n$. The {\it dihedral}
group $D_n$, order $2n$, is def\/ined as the {\it semidirect product}
$Z_n\odot Z_2$, where $Z_2$ performs the automorphism of going to
the inverse $g\rightarrow g^{-1}$ in $Z_n$: if $\alpha$ in $Z_2$,
$\alpha(g) = g^{-1}$, which in an Abelian group (as $Z_n$) is
automorphism ({\it anti}automorphism in a general group).

We already stated that $A$ Abelian and simple $\Longleftrightarrow A
= Z_p$, cyclic of prime order. A fundamental theorem written by \'E.
Galois his last night alive was:

\medskip

\noindent
{\bf Theorem (Galois, 1832).} {\it The alternative group ${\rm Alt}_n$ is
simple for $n
> 4$.}

\medskip

    We do not prove it. For lower degrees, we have the equivalences
\begin{gather*}
{\rm S}_1 = {\rm Id},\quad   {\rm Alt}_1 = \varnothing,\quad {\rm S}_2 = Z_2 ,\quad  {\rm Alt}_2 =
{\rm Id},\quad {\rm Alt}_3 = Z_3 ,\quad {\rm S}_3 = D_3 = Z_3 \odot Z_2 ,\\
{\rm Alt}_4 = V \odot Z_3 \quad  {\rm and} \quad   {\rm S}_4 = V \odot {\rm S}_3 .
\end{gather*}
Here $V = Z_2 \times  Z_2$ (direct product), the so-called {\it
Vierergruppe} of F.~Klein, with elements $(I,a,b,ab)$; $a$, $b$, $ab$
permutable under ${\rm S}_3$, so ${\rm Aut}(V) = {\rm S}_3$. $V$  is the f\/irst
non-cyclic group, as ${\rm S}_3$ is the f\/irst non-Abelian group.

$Q = \pm ({\rm Id}, i, j, k)$ is the {\it quaternion} group, of order 8,
where $i^2 = j^2 = (ij=-ji=k)^2 = -1$, etc. It is the f\/irst example of a
{\it dicyclic} group (see just below).

There is precisely an Abelian cyclic group $Z_n$ for each natural
number $n$. The three non-Abelian groups up to order eight are
${\rm S}_3=D_3$ (order 6), $Q$ and $D_4$ (order 8). Readers will be amused
to learn that there are nearly f\/ifty thousand million dif\/ferent
groups of order $2^{10}=1024$~\cite{Mil}, with ${\rm Part}(10) = 42$
Abelian: The number of Abelian groups of order $q = p^f$, power of a
prime, is ${\rm Part}(f)$; for example, there are {\it three} Abelian
groups of order $8 = 2^3$, to wit, $Z_8$, $Z_4 \times  Z_2$ and
$(Z_2)^3$; the later, $(Z_p)^f$,  are called {\it elementary Abelian groups}.

The importance of (f\/inite) simple groups lies in that they are the
{\bfseries \itshape atoms} in the category (of f\/inite groups), that is, any
f\/inite group is either simple or a particular extension of simple
groups; so it was a big advance when around 1980 mathematicians
realized (Gorenstein) that all {\it finite simple groups} were
already known.

A f\/inite group can be def\/ined by {\it generators and relations}; for
example, $\{ g; g^n=1\}$ describes~$Z_n$.  $\{g^n=\alpha^ 2=I,\,
\alpha\cdot g\cdot\alpha= g^{-1}\}$ describes $D_n$. $\{  g^{2m}
=\alpha^ 4=I,\, g^m =\alpha^ 2 ,\, \alpha \cdot
g\cdot\alpha^{-1}=g^{-1}\}$ describes the {\it dicyclic group},
$Q_m$ of order $4m$, so the previous $Q$ is just $Q_2$. Etc; see
\cite{Cox-1,Cox-2,Ram}, etc. The minimal number of generators is
called sometimes the {\it rank} of the (f\/inite) group.

\section{Families of f\/inite simple groups}\label{section2}

\subsection{Fields of numbers}

There are $(1+1+16)$ {\it families} of f\/inite simple groups, where
the f\/irst two are, as said, the cyclic Abelian group $Z_p$ (for any
prime number $p$), of order $p$, and the alternative (non-Abelian)
${\rm Alt}_n$ (for $n > 4$), of order $n!/2$. The 16 other families are
related to {\it finite geometries}, that is, they are
{\bfseries \itshape finite groups of Lie type}, or the analogous of matrix
groups over f\/inite f\/ields of numbers \cite{Car}. There are also 26
isolated f\/inite simple groups {\it not} in these families; they are
called {\bfseries \itshape sporadic groups} (Section~\ref{section3}).

Ordinary, continuous Lie groups are def\/ined as matrix groups with
real or complex entries: e.g.~${\rm SU}(5)$ is the set of $5 \times 5$
complex unitary matrices with $\det =1$, under matrix
multiplication. In fact, the f\/ields $\mathbb R$ and $\mathbb C$ are
the most conspicuous of the continuous f\/ields of numbers. There are
still two more extensions of the reals besides $\mathbb C$, yielding
``divison algebras'': $\mathbb H$ (the quaternions of Hamilton) and
$\mathbb O$ (the octonions of Graves); as neither are commutative
($\mathbb O$ is even not associative), they are not considered
f\/ields today ($\mathbb H$ is called a skew f\/ield), but have
plentiful applications; see e.g.~\cite{Con-1}.

Recall a {\bfseries \itshape field} $K$ is an algebraic structure with two
operations, sum and product; under the sum $K$ is an Abelian group,
with the neutral element written~0; under the product $K^*:=
K\setminus \{0\}$ is also and {\it Abelian} group, with the neutral
(unit) written~1. Equivalently, a f\/ield is a {\it commutative ring}
with all elements $k\ne 0$ {\it invertible}. Multiplication is
distributive, $k(j + m) = kj + km$; and $0\cdot k = 0$. The
characteristic of a f\/ield, ${\rm char}\, (K)$ is the minimum $n$ such
$1+1+\cdots  = n\cdot 1=0$: $\mathbb R$~or~$\mathbb C$ have {\it
char} zero (by def\/inition). For any f\/ield with ${\rm char}\,(K) = c$,
there is a~primitive f\/ield~$F_c$ with $c$ as characteristic,  so $K$ is a {\it field extension} of
$F_c$; for instance, the rational num\-bers~$Q$~$(=F_0)$ form the
primitive f\/ield of ${\rm char} =0$, with the reals $\mathbb R$ as a
(trascendent) extension, and the f\/ields $\mathbb F_p$ (see next) are
the primitive ones with ${\rm char} = p$. See \cite{Art}.

\subsection[Finite f\/ields $F_q$]{Finite f\/ields $\boldsymbol{{\mathbb F}_q}$}

Galois found also the {\it finite fields}, which we shall denote as
$\mathbb F_q$; here $q = p^f$ is an arbitrary {\it power} $f \ge 1$
of an arbitrary {\it prime} $p$; it is ${\rm char} (\mathbb F_{ p^f})= p$.
The notation $GF(q)$ for our ``Galois f\/ield'' $\mathbb F_q$ is also
very common.

We describe f\/irst the pure prime case, $f=1$: $\mathbb F_ p$ is a
set of $p$ elements $0, 1, \dots, p-1$ with sum and product {\it
defined} mod $p$; this establishes a f\/ield. For example, for the
smallest f\/ield $\mathbb F_2$, with only 1 and 0 as elements, the
rules are
\begin{gather*}
 0+0=0,\quad 0+1=1 , \quad 1+1 = 0,  \qquad        0\times 0=0 ,\quad  0\times 1 = 0,\quad  1\times 1
 =1.
\end{gather*}

These rules for $\mathbb F_2$ extend easily to any prime $p$, so
$\mathbb F_ p$ makes now sense as a f\/inite f\/ield of~$p$ elements:
sum and product mod $p$. For $q = p^f$, $f >1$ arbitrary, the f\/ield
structure in $\mathbb F_q$ is dif\/ferent: consider the elementary
Abelian group $(Z_p)^f := Z_p \oplus  Z_p \oplus Z_p \oplus\cdots
\oplus Z_p$, $f$ times: the sum in $\mathbb F_p$ is def\/ined as in
this group $(Z_p)^f$. But the product in ${\mathbb F}_q^* \equiv
\mathbb F_q\setminus \{0\}$ is def\/ined as in the cyclic group
$Z_{q-1}$; one checks this def\/ines a {\it bona fide} commutative
f\/ield. So $ \mathbb F_2$, $\mathbb F_3$, $\mathbb F_4$, $\mathbb F_5$,
$\mathbb F_7$, $\mathbb F_8$, $\mathbb F_9$ are all f\/inite f\/ields $K$ with
$|K| \le 10$.

We exemplify the f\/ield law for the case $q=2^2$, or $\mathbb F_4$:
it has four elements, $0$, $1$, $\omega$, $\overline\omega$: the last
three form $Z_3$ as multiplicative group, as said;
$\omega:=\exp(2\pi i)/3)$. The rules are
\begin{gather*}
\text{the sum is} \ \   1+1=0 ,\quad 1+ \omega=\overline\omega, \quad \overline\omega+\omega=1 ,\quad \omega+\omega=0
 ,\quad  \overline\omega+\overline\omega=0 ,\quad {\rm etc.}, \nonumber
 \\
\text{the product}  \ \ 0\cdot({\rm any})=0,\quad  1\cdot ({\rm any})
= ({\rm any}),\quad \omega\cdot\omega= \overline\omega,\quad  \omega\cdot
\overline\omega=1,  \quad {\rm  etc}. 
\end{gather*}

Now, we construct {\it finite geometries} over these f\/inite f\/ields.
A vector space $V$ over $\mathbb F_q$ will have a f\/inite dimension,
say $n$; it will have $q^n$ elements, and can be written (compare
$\mathbb R^n$ or $\mathbb C^n$) as~$\mathbb F_q^n$. Matrices
(=~linear maps) with entries on $\mathbb F_q$ will be endomorphisms
of these vector spaces, indeed ${\rm End}(\mathbb F_q^n)$ will have
$q^n\times q^n$ elements. We are interested in invertible matrices,
which will form a (non-commutative for $n>1$!) group under
multiplication; let us call ${\rm GL}(V) = {\rm GL}_n(q)$ the group of
invertible matrices of dim $n$ over $\mathbb F_q$. For $n=1$ we have
$|{\rm GL}_1(q)|=|{\mathbb F}_q^*| = q-1$, as the zero is to be excluded.

Let us look now at the order of ${\rm GL}_2(q)$; it is
\begin{gather*}
 |{\rm GL}_2(q)| = \big(q^2-1\big)\big(q^2-q\big) = (q+1)\cdot q\cdot (q-1)^2 ,
\end{gather*}
because in the f\/irst row, all elements cannot be 0, and the second
row must be linearly independent from the f\/irst. Notice the order
$|{\rm GL}_2|$ is divisible by~6, and if $q$ odd, by~48.

A simple generalization gives the order of ${\rm GL}_n(q)$:
\begin{gather*}
|{\rm GL}_n(q)|= \big(q^n-1\big)\big(q^n-q\big)\big(q^n-q^2\big)\cdots \big(q^n -q^{n-1}\big) .
\end{gather*}

\subsection{Projective spaces}

$\mathbb F_qP^k$ is the set of one-dim subspaces in $V=\mathbb
F_q^{k+1}$, as $\mathbb RP^n$ is the real $n$-dim projective space.
The projective line $\mathbb F_qP^1$ is the set of lines in $\mathbb
F_q^2$, with $(q^2-1)/(q-1) = q+1$ points; the projective plane
$\mathbb F_qP^2$ are lines in $\mathbb F_q^3$, with $(q^3-1)/(q-1)
= q^2 + q + 1$ points (and lines): in general,
\begin{gather}\label{10}
 | \mathbb F_qP^k | = \big(q^{k+1} - 1\big)/(q -1) = q^k + q^{k-1} + \cdots+ q +1 \quad {\rm
 points} .
\end{gather}

A theorem of Wedderburn (1905) assures that any f\/inite ``f\/ield'' is
commutative; one shows also \cite{Car} that these $\mathbb F_q$,
with $q = p^f$ exhaust all {\it finite} f\/ields. The f\/inite groups
 over the vector spaces over the f\/inite f\/ields are called also {\it Chevalley groups}, as C.
Chevalley proved (1955) that the all the conventional continuous Lie
groups over $\mathbb R$, $\mathbb C$, have analogous over $\mathbb
F_q$.

Now ${\rm GL}_n(q)$ is not simple, as the ``diagonal'' entries $\approx
F_q^*$ form a normal Abelian subgroup (the center); also the natural
map $\det: {\rm GL}\rightarrow \mathbb F_q^*$ has a~kernel: the unimodular
group ${\rm SL} = {\rm SL}_n(q)$
\begin{gather*}
 {\rm GL}_n(q)/ \mathbb F_q^* := {\rm PGL}_n(q),  \qquad      {\rm SL}_n(q)\rightarrow  {\rm GL}_n(q)\rightarrow
{\mathbb F}_q^* .
\end{gather*}

Now ${\rm SL}_n$ might still have central elements, those $k$ in ${\mathbb F}_q$
with $k^n = I$. Denote ${\rm PSL}_n(q)$ the quotient by these possible
central elements, ${\rm PSL}_n(q) = {\rm SL}_n(q)/({\rm Center})$. There is now the
capital theorem of Jordan--Dickson (ca.~1900):

\medskip

\noindent
{\bf Theorem.} {\it ${\rm PSL}_n(q)$    is simple, except $n=2$ and $q= 2$ or~$3$.}

\medskip

We shall not try to prove this \cite{Dieu}, but note only that
${\rm PSL}_2(2) = {\rm GL}_2(2)\equiv {\rm S}_3$, and ${\rm PSL}_2(3)={\rm Alt}_4$. Both ${\rm S}_3$ and
${\rm Alt}_4$ are solvable groups, as it is ${\rm S}_4$.

This gives the third (and most important) {\it doubly} parametric
family of {\it finite simple groups}. The order is easy to f\/ind:
\begin{gather*}
|{\rm PSL}_n(q)| = \prod _{m=1}^{m=n-1} \big(\big(q^n - q^m\big)/(q-1)\big)/ \{n, q-1\} ,
\end{gather*}
where  $\{n, q-1\}$   indicates the order of the center of
${\rm SL}_n(q)$. The complete diagram is
\begin{gather*}
\begin{array}{@{}ccccc}
           \{n,q-1\} & \rightarrow & {\rm SL}_n(q) & \rightarrow & {\rm PSL}_n(q) \\
           \downarrow & \, & \downarrow & \, & \downarrow \\
        {\mathbb F}^*_q & \rightarrow & {\rm GL}_n(q) & \rightarrow & {\rm PGL}_n(q) \\
           \downarrow & \, & \downarrow & \, & \downarrow \\
           {\mathbb F} & \rightarrow & {\mathbb F}_q^* & \rightarrow &  \{n,q-1\}
         \end{array}
\end{gather*}
where ${\mathbb F}:={\mathbb F}^*_q/\{n-1,q\}$.

\subsection{Other (three) bi-parametric families}

The development now follows closely that of matrix groups over
$\mathbb R$ or $\mathbb C$: namely one looks for the invariance
group (stabilizer) of some tensorial objects: a quadratic form $Q$
(to def\/ine the orthogonal group ${\rm O}$), a 2-form $\omega$  (for the
symplectic group ${\rm Sp}$), besides a volume form $\tau$  (whose
stabilizer is the unimodular ${\rm SL}$) and studies the subquotients
(quotients of subgroups) which are simple; we omit the details (for
example, ${\rm O}(n)$ might admit signature, as in ${\rm O}(p,q; \mathbb R)$;
${\rm SO}$~might still be not simple, but the {\it commutator} $\Omega$
should be; ${\rm Sp}$ exists only in even dimensions, etc.), only to signal
two other biparametric families of f\/inite simple groups as
\cite{Car,Art}{\samepage
\begin{gather*}
 {\rm P \Omega}_n(q)\ \ \text{(not standard notation)} \quad      {\rm and}  \quad
 {\rm PSp}_n(q).
\end{gather*}
We shall also omit the orders, which can be checked in many places
\cite{Car,Dieu}.}

For Lie groups, we have also the {\it unitary} (sub)groups. Do they
appear here? Yes! Recall the {\it conjugation} automorphism of the
complex f\/ield $\mathbb C$, $z = x+iy  \rightarrow  \overline z = x -
iy$ (the real f\/ield $\mathbb R$ has no f\/ield automorphisms $\ne 1$).
One shows that \cite{Art}:
\begin{gather*}
\text{the prime f\/ields}\ \ \mathbb F_p\ \ \text{have not f\/ield automorphisms} \ (\ne~{\rm Id}), \label{16}
\\\text{the f\/ields of order}\ \  q = p^f,\ \  f > 1,\ \  {\rm have}\ \ {\rm Aut}(\mathbb F_{q}) = Z_f   
\ \ \text{in particular}\ \ {\rm Aut} (\mathbb F_{p^2})= Z_2.
\end{gather*}

If $F=\mathbb F_4 \equiv\{  0, 1,\omega  ,\overline\omega\}$, the
${\rm Aut} \ne I$ is $\omega \longleftrightarrow \overline\omega$.

Field automorphisms allow {\it semilinear} applications $s: V
\rightarrow  V$:
\begin{gather*}
   s(x+y) = s(x) + s(y), \ \   {\rm but} \ \    s(\lambda x) =\lambda^\sigma
   s(x),
\end{gather*}
where $\sigma :\lambda\rightarrow \lambda^\sigma$       is the f\/ield
automorphism, $\lambda\in \mathbb F_q$, $\sigma\in {\rm Aut}(\mathbb
F_q)$; in particular ${\rm \Gamma L}_n(q)$ is the group of {\it all}
invertible $n\times n$ semilinear maps in $\mathbb F_q^n$. With
involutory f\/ield automorphisms, always present for $q=p^f$, $f >1$
even, we form the hermitian sesquilinear product, $h(x, y) \equiv x
\cdot y$. Now the unitary group is def\/ined like in the complex case,
as the stabilizer of the hermitian form among {\it linear} maps:
\begin{gather*}
  U(h):\ \{  U \in  {\rm GL}, \  h(x, y) = h(Ux, Uy)\}.
\end{gather*}

So one obtains thus the {\it fourth} family of f\/inite simple groups
of Lie type by considering the pertinent ${\rm PU}$ group; we omit the
details~\cite{Car}. {\it In toto} we have four doubly-parametric
families of f\/inite matrix simple groups (for ${\rm O}$ and ${\rm U}$ the
notation is not standard):
\begin{gather*}
{\rm PSL}_n(q),\quad   {\rm P\Omega}_n(q), \quad {\rm PSp}_n(q) \quad {\rm
and}\quad {\rm PSU}_n(q).
\end{gather*}

In particular, one has the following short catalogue of the
{\bfseries\itshape six} non-Abelian f\/inite simple group or order $|G| <
2000$:
\begin{gather*}
\begin{array}{@{}c@{\,\,\,}c@{\,\,\,}c@{\,\,\,}c@{\,\,\,}c@{\,\,\,}c@{\,\,\,}c@{}}
  & {\rm Alt}_5 & {\rm PSL}_3(2)= {\rm PSL}_2(7) & {\rm Alt}_6 ={\rm PSL}_2(9) & {\rm PSL}_2(8) & {\rm PSL}_2(11) & {\rm PSL}_2(13) \\[1ex]
{\rm\bf order} & 60 & 168 & 360 & 504 & 660  & 1092
       \end{array}
\end{gather*}

Other authors follow Cartan's classif\/ication, writing $A_n(q)$,
$B_n(q)$, $C_n(q)$ and $D_n(q)$: $A$ stands for ${\rm SL}$ or ${\rm U}$, $C$ for
symplectic, $B$ for odd orthogonal, $D$ for even orthogonal.

\subsection{Finite simple groups of exceptional-Lie type}

Chevalley realized (1955) that also {\it all} the exceptional Lie
groups, namely $G_2$, $F_4$ and the $E_{6,7,8}$ series could also be
constructed in f\/inite geometries; for example, $G_2$ is the
stabilizer of a 3-form in a 7-dim space~\cite{Boy-1}, and the
construction works for {\it any field} $K$, including f\/inite f\/ields.
So one builds the next f\/ive {\it uniparametric} families of f\/inite
(mainly) simple groups, denoted
\begin{gather}\label{23}
   G_2 (q),\quad  F_4(q),\quad  E_6(q),\quad  E_7(q)\quad  {\rm and}\quad
   E_8(q).
\end{gather}

For example, the order of the smallest (casually {\it not} simple
for $q=2$) is
\begin{gather*}
|G_2(q)| = q^6 (q^6-1)(q^2-1) \qquad   ({\rm e.g.} = 12 096\ \
{\rm for}\  q=2).
\end{gather*}

There are {\it two further sets of families} of f\/inite simple
groups of Lie type. Recall, f\/irst, that the (continuous) families of
simply connected compact Lie groups ${\rm SU}(n)$, ${\rm Spin}(2n)$, plus the
isolated cases ${\rm O}(8)$ (${\rm Spin}(8)$) and $E_6$, all have outer
automorphisms: one can read them of\/f from the Dynkin diagram; for
example, in ${\rm SU}(n+1)\equiv A_n$: if $n >1$, one interchanges the
$k$-th node with the $(n-k)$-th; similarly for $E_6$; and in ${\rm O}(8)
\equiv D_4$ there is {\it triality}, i.e., permutation of any of the
three outer nodes. In the f\/inite f\/ield case, twisting by these
automorphisms produces {\it four} new families (Steinberg) (there
are some restrictions on dimensions and on the f\/ields $\mathbb F_q$,
that we omit), usually written as
\begin{gather*}
 {}^2A_n(q)\  (n > 1),\quad   {}^2D_n(q) \ (n\ne 4),\quad   {}^2E_6(q)\quad   {\rm
 and}\quad
 {}^3D_4(q).
\end{gather*}

Finally, for the three cases $B_2$, $G_2$ and $F_4$, with Dynkin
diagrams
\begin{gather*}
   B_2: \ \circ \Longrightarrow \bullet \qquad G_2: \ \circ \Rrightarrow
   \bullet  \qquad F_4: \
   \circ---\circ\Longrightarrow\bullet---\bullet
\end{gather*}
reversing the arrow the Lie groups are the same, but in the f\/inite
f\/ield case we get in most cases (Suzuki, Rae) {\it three} new
families of f\/inite simple groups, written also as
\begin{gather*}
 {}^2B_2(q), \quad   {}^2G_2(q) \quad   {\rm and } \quad    {}^2F_4(q)
\end{gather*}
(with some restrictions again on the f\/ields).

So in total we have the 1+1+16 (= (4)+(5)+(4)+(3)) = 18 (bi- or
mono- parametric) {\it families} of f\/inite simple groups, completed
around 1960:
\begin{alignat*}{4}
 & Z_p, \ {\rm Alt}_{n>4}   && 2 && \text{Mono-}p \hskip2cm & \\
& {\rm PSL}_n(q), \ {\rm P\Omega}_n(q), \ {\rm PSp}_n(q), \ {\rm PU}_n(q)   && 4 && \text{Bi-}p \hskip2cm & \\
& G_2(q), \ F_4(q), \ E_6(q), \ E_7(q), \ E_8(q)   && 5 &&  \text{Mono-}p \hskip2cm& \\
& {}^2A_n(q)\  (n > 1), \ {}^2D_n(q), \ {}^2E_6(q)\  {\rm and}\
 {}^3D_4(q)  \qquad \quad && 2\,{\rm Bi}\ \ \&  \ \ && \text{2\ Mono-}p \hskip2cm& \\
& {}^2B_2(q), \ {}^2G_2(q), \ {}^2F_4(q) \hfill && 3 && \text{Mono-}p
\hskip2cm& \\[-1ex]
 &&&   {---} && \hskip2cm& \\[-1ex]
  &&&    18  && \hskip4cm&
\end{alignat*}

It is a not totally understood fact that the order of any
(non-Abelian) f\/inite simple group is divisible by 12 (divisibility
by 2 was proven in 1963: it is the famous Feit--Thomson theorem).
Burnside already proved that $G$ f\/inite simple $\Longrightarrow |G|$
divisible by three dif\/ferent primes; the smallest possibility is already 	realized, as $|{\rm Alt}_5| = 60 = 2^2\cdot 3\cdot 5$.

\section{Sporadic groups}\label{section3}

\subsection{Higher transitivity}\label{section3.1}

Let $G \circ \rightarrow  X$ be an action of group $G$ in space $X$.
The action is transitive (as said) if $G(x)=X$, i.e.~any point $y$
is reachable from any other $x$ through some $g$ in $G$ (so $g\cdot
x=y$): there is only an orbit. If $G_x$ is the stabilizer of point
$x$, it acts in $X \setminus \{x\}$ in a natural way; if this second
action is also {\it transitive}, we say that $G$ acts 2-transitive
in $X$; it is the same as saying: any two points $y\ne z$ can go to
any other distinct points $y'$, $z'$. Let $G_{xy}\subset G_x$ be the
stabilizer of the second action: again this group acts in
$X\setminus\{ x, y\}$ naturally; if this action is again transitive,
we say that $G$ acts 3-{\it transitive} in $X$; also, this is
equivalent to say: any three distinct points~$u$,~$v$,~$w$ can go to any
other three distinct~$u'$,~$v'$,~$w'$ by the action of some $g \in  G$.
In this form one speaks of $k$-transitive action of a group $G$ in a
space $X$. Finally, one says that the action $G \circ\rightarrow X$
is \textit{sharp} $k$-transitive, if the last stabilizer is just $I={\rm Id}$.
1-transitive actions are called just transitive; sharp 1-transitive
actions are called {\it regular}. If e.g.~$G$ acts sharp
3-transitive in~$X$, one has naturally $|G|=(|X|)·(|X|-1)·(|X|-2)$;
etc.

Actions more than transitive (=~1-transitive) are rare. For the
common example,  ${\rm SO}(n+1)$ acts transitive on the sphere $S^n$, but
the little group ${\rm SO}(n)$ f\/ixes two (antipodal) points and describes
the parallels with the f\/ixed points as poles (say, $N$ and $S$): the
action is 1-transitive with stabilizer $\ne I$ (so not {\it sharp}
1-trans), but not 2-transitive. Also, with ${\rm GL}_n(\mathbb R)$ acting
on $V\setminus \{ 0\}$, the action is transitive; but if $g\cdot
x=x$, the stabilizer $G_x$ leaves also pointwise f\/ixed the ray of
$x$, $\{x\}$, so the action again is not 2-transitive.

The paradigmatic example of sharp $n$-transitive action is, of
course, the symmetric group~${\rm S}_n$ acting naturally sharp
$n$-transitive in $n$ symbols: it is the very def\/inition of~${\rm S}_n$.
However, the reader should check that ${\rm Alt}_n$ is only sharp ($n-2$)
transitive in these $n$ symbols (the clue is that ${\rm Alt}_3 \equiv
Z_3$). One sees that the action of ${\rm S}_n$ (or ${\rm Alt}_n$) on $n$
symbols is {\bfseries\itshape sharp $n$-transitive}, (resp.
sharp-($n-2$)-{\it trans}) because the little group of the last
action is the identity.

Apart from ${\rm S}_n$, and ${\rm Alt}_n$, actions more than 3-transitive are
very exceptional. But we are just to show, as another example, a
whole family of {\it natural generic sharp} 3-transitive actions.

Consider again $\mathbb F_q$, the Galois f\/ield, and the projective
line $\mathbb F_qP^1$, or the set of 1-dim subspaces or lines
through origin in $\mathbb F_q^2$; it has $q+1$ element, with
$\infty$ added (recall the real projective line $\mathbb RP^1 \equiv
S^1$, is the circle, as one-point compactif\/ication of the line
$\mathbb R$, $\mathbb RP^1 \equiv \mathbb R \cup \{\infty\}$). As we
have added the ``point at inf\/inity'', one can put the action in the
``homographic form'', i.e.\ $x \rightarrow (ax+b)/(cx+d)$ with the
$\det = (ad-bc) \ne 0$. This def\/ines an ef\/fective action of
${\rm PGL}_2(q)$ on $\mathbb F_qP^1$. The action is transitive, with
little group of $\infty: (abcd) = (ab0d)$ with $ad\ne 0$. The new
action is now {\it affine}, $x\rightarrow (a/d)x + b/d$, $a\ne 0\ne
d$; it is still {\it trans} on the very f\/ield $\mathbb F_q$ (as
1-dim vector space), $V = \mathbb F_q$. The stabilizer of the origin
0 is now $(b/d)=0$ or $b=0$, and the remaining group are the
dilatations $x \rightarrow\lambda x$, $\lambda\ne 0$. This is again
transitive in $F^*_q = F_q\setminus \{ 0\}$, but in this third
action there is no leftover little group (but ${\rm Id}$): hence the last
action is sharp:
\begin{enumerate}\itemsep=0pt
\item[] ``The action of ${\rm PGL}_2(q)$ by homographies on the projective line
$\mathbb F_qP^1$ is sharp 3-transitive''.
\end{enumerate}

In fact, this result {\it holds for any field} $K$. Notice that the
unimodular restriction group ${\rm PSL}_2(q)$ is only 2-transitive,
non-sharp. One checks also that ${\rm PGL}_{n+1>2}(q)$ is not 3-transitive
in the corresponding projective space $\mathbb F_qP^n$. The case
${\rm PGL}_3(4)$ will occupy us later.

If an action (of $G$ on $X$) is sharp 1-transitive, clearly $|G| =
|X|$, as said. For the projective line of above, the action being
sharp 3-transitive, we have $|{\rm PGL}_2(q)| = (q+1)\cdot (q)\cdot
(q-1)$, divisible at least by 6, and, if $q$ is odd, at least 24.

\subsection{The f\/irst level of sporadic groups: the f\/ive Mathieu groups}\label{section3.2}

It turns out that the sporadic groups come in four classes, {\it
three} consecutive levels {\it plus} the Pariahs. The levels are
Mathieu's (5 groups), Leech's (7 groups) and Monster's (8), plus 6
Pariah groups ({\it families} instead {\it levels} is also common
name). So there are {\bfseries\itshape 26 sporadic groups}.

Already from 1861 \'E. Mathieu, a notable French mathematical
physicist (Mathieu equation, Mathieu functions, etc.) found {\it
five} f\/inite simple groups not in the above families; by def\/inition,
these groups will be called (as said) {\it sporadic} (Burnside); so
sporadic groups are f\/inite, simple groups, and not in the above 18
families. We shall describe the Mathieu groups as constituting the
{\bfseries \itshape first level of sporadic groups}. To introduce them
properly we shall use the concept of multiple transitivity of
previous Section~\ref{section3.1} \cite{Con-2,Grie}. These groups are named
${\rm M}_{11}$, ${\rm M}_{12}$, ${\rm M}_{22}$, ${\rm M}_{23}$ and ${\rm M}_{24}$.

We gave f\/irst the order of the f\/irst two: ${\rm M}_{11}$ is a sharp
4-transitive (simple) group acting in 11 symbols. Its order is
therefore
\begin{gather*}
   |{\rm M}_{11}| = 11\cdot 10\cdot 9\cdot 8 = 7920.
\end{gather*}

In fact, Mathieu was looking for higher-than-3-transitive actions;
the {\it simplicity} of ${\rm M}_{11}$ was proved later. Mathieu also found
that there is a natural {\it $12$-ampliation} to ${\rm M}_{12}$: this new
group is sharp 5-transitive in 12 symbols, so its order is
\begin{gather*}
 |{\rm M}_{12}| = 12\cdot |{\rm M}_{11}| = 95 040.
\end{gather*}

Indeed, as the reader may expect, ${\rm M}_{11}$ is the f\/irst stabilizer
of ${\rm M}_{12}$, with {\it sharp $5$-trans action} in 12 objects; for {\it
ampliations} (in this sense) see \cite{Why}.

We present a view how ${\rm M}_{11}$ and ${\rm M}_{12}$ are really constructed
\cite{Boy-2}. Consider the groups ${\rm Alt}_6$ and ${\rm PSL}_2(9)$. Both are
of the same order and (of course) simple:
\begin{gather*}
 |{\rm Alt}_6| = 6\cdot 5\cdot 4\cdot 3= 360, \qquad   |{\rm PSL}_2(9)| = (9^2-1)(9^2-9)/8/2 = 10 \cdot 9 \cdot 8/2 =
 360.
\end{gather*}

It turns out that both groups are {\it isomorphic}. Now, the natural
2-extension of ${\rm Alt}_n$ is ${\rm Alt}\cdot 2= {\rm Sym}_n$, and the 2-natural one
of ${\rm PSL}_2(q)$ is to ${\rm PGL}_2(q)$ (see diagram below). It turns out
that ${\rm Sym}_6$ and ${\rm PGL}_2(9)$ are of course of same order 720, {\it
but not isomorphic}. Hence, as both come from ${\rm Alt}_6$, there must be
at least two classes of {\it outer} automorphisms in ${\rm Alt}_6$~\cite{Green}, say~$\alpha$~and~$\beta$ to generate dif\/ferent groups. But
then, as $\alpha$ and $\beta$ are involutory and commute,
$\alpha\beta$ is also another involutory {\it outo}, and together
they form $(1,\alpha ,\beta  ,\alpha\beta )$, the {\it Vierergruppe}~$V$ of Klein (def\/ined above):
\begin{gather*}
 {\rm Out}({\rm Alt}_6) = V.
\end{gather*}

Let us call ${\rm M}_{10}$ the 2-extension from ${\rm Alt}_6$ due to
$\alpha\beta$. We have the following diagram
\begin{gather*}
\begin{array}{@{}cccccccc}
              & \alpha:     & \  & {\rm Alt}_6    & \longrightarrow & {\rm Sym}_6          & \,          &                          \\[1ex]
              &             &       & \|             &             & \not\hspace{-1.5pt}\|    & \searrow    &                     \\[1ex]
              & \beta:      & \  & {\rm PSL}_2(9) & \longrightarrow & {\rm PGL}_2(9) & \longrightarrow & {\rm P\Gamma L}_2(9) \\[1ex]
              &             &       &                & \searrow    &                      & \nearrow    &                           \\[1ex]
              &\alpha\beta: & \  &                &             & {\rm M}_{10}         &             &                           \\[1.5ex]
  {\bf order} &             & \ & 360            &             & 720                  &             & 1440
\end{array}\qquad \qquad {\bf Diagram~II}
\end{gather*}

There is an extra outer automorphism in the three groups of the
middle column, the missing one, whose extension goes in the three
cases to the same group, ${\rm P\Gamma L}_2(9)$ of order 1440. Here
${\rm P\Gamma L}$ (projective semilinear) refers to the
f\/ield involutory automorphism of $F_9$, as $9 = q^f = 3^2$ and 3
prime, see~\cite{Car}. This {\it extra} external automorphism of ${\rm Sym}_6$
was already noticed by Sylvester in 1849 \cite{Cox-3}, as ${\rm S}_6$ is
the only symmetric group ${\rm S}_n$ with an outer automorphism. For a
related construction with graphs see also \cite{Cox-3}.

The full transitivity chain is now ($Q$ is the quaternion group, see
above)
\begin{gather*}
 \begin{array}{@{}c@{\,\,\,}c@{\,\,\,}c@{\,\,\,}c@{\,\,\,}c@{\,\,\,}c@{\,\,\,}c@{\,\,\,}c@{\,\,\,}c@{\,\,\,}c@{}}
{\rm M}_{12} & \supset & {\rm M}_{11} & \supset & {\rm M}_{10} & \supset & {\rm M}_9\approx
Z_9\odot{\mathbb Q} & \supset & {\rm M}_8\approx
{\mathbb Q} & \, \\[1ex]
\text{sh 5-trans} & \, & \text{sh 4-trans} & \, & \text{sh 3-trans} & \, & \text{sh 2-trans} & \,
& \text{sh 1-trans=regular} & \, \\[1ex]
95 040  & \, & 7 920 & \, & 720 & \, & 72 & \, & 8 & {\rm\bf  order}
 \end{array}
\end{gather*}

There are {\it other three Mathieu} groups, also simple and multiple
transitive (but {\it not} sharp), studied by Mathieu himself between
1861 and 1873. We start again from the known groups ${\rm Sym}_8$ and
${\rm GL}_3(4)$, and look at two related groups of the same order, viz.:
\begin{gather*}
 |{\rm Alt}_8| = 8\cdot 7\cdot 6\cdot 5\cdot 4\cdot 3= 20 160 = |{\rm PSL}_3(4)| =\big(4^3-1\big)\big(4^3- 4\big)\big(4^3-
 4^2\big)/3/3 .
\end{gather*}

It turns out that these groups are {\it not} isomorphic (but
${\rm GL}_4(2)= {\rm PSL}_4(2)$ is of the same order, and isomorphic with
${\rm Alt}_8$), although they are still (of course) {\it simple}. The
second group ${\rm PSL}_3(4)$ has a natural action in the projective {\it
plane} $\mathbb F_4P^2$ with $(q^3-1)/(q-1) = q^2 + q +q|_{q=4} =
21$ elements (\ref{10}), with a natural (nonsharp!) 2-transitive
action: indeed, from above
\begin{gather*}
 |{\rm PSL}_3(4)| = 21\cdot 20\cdot 48.
\end{gather*}

We can call it also ${\rm M}_{21}$, as it admits a natural 22-ampliation
to ${\rm M}_{22}$, 3-trans in 22 and simple!! (see \cite{Why}). On its
turn, there are {\it two} more ampliations, to ${\rm M}_{23}$ and to
${\rm M}_{24}$, with the full chain again (we do not specify ${\rm M}$, only
their orders):
\begin{gather*}
\begin{array}{@{}ccccccccc@{}}
{\rm M}_{24} &\supset & {\rm M}_{23} & \supset & {\rm M}_{22} & \supset & {\rm M}_{21} & \supset & {\rm M}_{20} \\
\text{5-trans} & \; &  \text{4-trans} & \; &  \text{3-trans} & \, &  \text{2-trans} & \; &  \text{1-trans}
 \\[1ex]
24\cdot 23\cdot 22\cdot 21\cdot 20\cdot 48    & \;& \; & \; & \, &
\,& 21\cdot 20\cdot 48 & \; & 20\cdot 48
\end{array}
\end{gather*}

The f\/irst thorough study of these Mathieu groups is by Witt (1938);
see~\cite{Con-2} and~\cite{Grie}. The Mathieu group ${\rm M}_{24}$ in
particular is very important, being related to several groups of the
next two levels of sporadic groups, and having lately being used also in
physics~\cite{Egu}; it also contains the other four Mathieu groups
as subgroups.

\subsection{The second level of sporadic groups (Leech lattice, Conway groups)}

Around 1960 interest on f\/inite simple groups increased, as Janko in
Australia found (1965) the f\/irst new sporadic group (${\rm J}_1$) a
century later after Mathieu's. It turns out that ${\rm J}_1$ had order 175560, and it is a {\it Pariah} group (see below). The trigger for the
second level of sporadic groups was a particular lattice discovered
by J.~Leech, which lives in 24 dimensions (a {\bfseries\itshape $m$-dim
lattice} is the $Z$-span of a vector base in $\mathbb R^m$ space,
here $m = 24$). If $\{e_i\}$ are $m$ linearly independent vectors in
$\mathbb R^m$, the points $x=\sum_{ i=1}^{24} n_i\, e_i$ ($n_i\in
\mathbb Z$) form a lattice. The {\it Leech lattice} $\Lambda_{24}$
was found by J.~Leech when working on coded message transmission; it
optimizes by far the best sphere packing in the dimension, with 196560 spheres touching a central one (kissing number~\cite{Con-2}),
and has other curious properties. The number of edges of length four
(none of length two) in the unit Leech lattice is also 196560, a
number related both to the sphere packing and to the Monster group,
see below. The number $24 = 2^3\cdot 3$ no doubt has remarkable
properties.

John H.~Conway found in 1968 the automorphism group of the Leech
lattice (isometries f\/ixing the center), an enormous group of order
\begin{gather*}
|{\rm Aut}(\Lambda_{ 24})| \equiv | {\rm Co}_0| = 2^{22}\cdot 3^9\cdot 5^4\cdot
7^2\cdot 11\cdot 13\cdot 23 \approx 8.31\cdot 10^{18}.
\end{gather*}

This ``zero degree'' Conway group is {\it not} simple, but it has a
{\it simple quotient}, named ${\rm Co}_1= {\rm Co}_0/Z_2$. Two other simple
groups were discovered by Conway also, ${\rm Co}_2$ and ${\rm Co}_3$, as
dif\/ferent stabilizers. In rapid succession {\it four} more simple
groups were discovered, all related to the Conway groups. They are
called Higman--Sims (HS), MacLaughlin (McL) Hall--Janko (HJ, also
called ${\rm J}_2$) and Suzuki (Suz). See e.g.~\cite{Con-2,Grie}, for the
total of {\it seven} groups related to the Leech lattice.

We do not elaborate, but include here just the orders (in prime
factors) of the {\bfseries \itshape seven second-level sporadic
groups}:
\begin{table}[h!]\centering
\begin{tabular}{cc}
Group & Order \bsep{1pt} \\
\hline
\tsep{3pt}Conway$_1$, Co$_1$ &  $2^{21}\cdot 3^9\cdot 5^4\cdot 7^2\cdot
11\cdot 13\cdot 23$\\
Conway$_2$, Co$_2$  &  $2^{18}\cdot 3^6\cdot 5^3\cdot 7\cdot 11\cdot 23$\\
Conway$_3$, Co$_3$   &       $2^{10}\cdot 3^7\cdot
5^3\cdot 7\cdot 11\cdot 23$ \\
Higman-Sims, HS &   $2^{9}\cdot 3^2\cdot 5^3\cdot 7\cdot 11$ \\
MacLaughlin, McL &  $2^{7}\cdot 3^6\cdot 5^3\cdot 7\cdot 11$ \\
Hall--Janko or ${\rm J}_2$, HJ  &   $2^{7}\cdot 3^3\cdot 5^2\cdot 7$\\
Suzuki,   Suz  &    $2^{13}\cdot 3^7\cdot 5^2\cdot 7$
\end{tabular}
\end{table}

\subsection{The third level: the Monster group}

Fischer and Griess independently suspected around 1973 the existence
of a very large sporadic (=~isolated f\/inite simple) group, called the
{\it Monster group} $\mathbb M$ (other names were $F_1$,
Fischer--Griess group, or friendly giant). The group was f\/inally
constructed by Griess in 1980 as an automorphism group of a remarkable
commutative but non-associative algebra with 196884 dimensions! (we
omit details, see e.g.~\cite{Gan} or \cite{Ron}), and it is of order
\begin{gather}\label{41}
      | \mathbb M | = 2^{46}\cdot 3^{20}\cdot 5^9\cdot 7^6\cdot 11^2\cdot
13^3\cdot 17\cdot 19\cdot 23\cdot 29 \cdot 31\cdot 41\cdot 47\cdot 59\cdot 71
\approx 8.08\cdot 10^{54}.
\end{gather}

In an gram atom there are $\approx 10^{24}$ atoms: the order of the
Monster group is close to the number of atoms in planet Jupiter! A
new modern building of the Monster group is via the {\it vertex
operators}, a specif\/ic construct in super string theory that we omit~\cite{Fre}. (The Frenkel et al.\ construction (1984) amounts to a
true quantum-mechanical build-up of the Monster group). R.~Borcherds
took the theory one step further (1988) by generalizing the
so-called Kac--Moody algebras.

The Monster $\mathbb M$ is by far the largest of the sporadic
groups, is related to all but four or six (depending on counting) of
the 26 sporadic groups, constituting the top of the {\it third
level} of sporadic groups. Fischer was also responsible for the next
largest group, the baby Monster~${\rm B}$ ($|{\rm B}| \approx 4\cdot 10^{33}$)
as well for another triplet of sporadics (${\rm Fi}_{22}$, ${\rm Fi}_{23}$ and
${\rm Fi}_{24}$) related to the analogous in the second Mathieu series
${\rm M}_{22\text{-}23\text{-}24}$, but much larger; all are subgroups of $\mathbb M$. A~few more groups, related also to $\mathbb M$, were constructed to
complete the {\it third level} of sporadic groups. We omit details,
limiting ourselves to exhibiting the names and some data of the
remaining 8~sporadic groups in the third level. Here is the
table with the {\bfseries \itshape third level} of the 8 sporadic
(simple) groups:
\begin{table}[h!!]\centering
\begin{tabular}{cc}
Group & Order \bsep{1pt} \\
\hline
\tsep{3pt}
Monster, $|\mathbb M|$    &        $\approx 8\cdot 10^{54}$\bsep{1pt}\\
baby Monster, B    &      $\approx 4\cdot 10^{33}$ \bsep{1pt}\\
Fischer$_{24}$    &         $\approx 1\cdot 10^{24}$ \bsep{1pt}\\
Fischer$_{23}$    &          $\approx 4\cdot 10^{18}$ \bsep{1pt}\\
Fischer$_{22}$   &           $\approx 64\cdot 10^{12}$ \bsep{1pt}\\
Harada--Norton, HN  &           $\approx 2\cdot 10^{14}$ \bsep{1pt}\\
Thomson, Th  &         $\approx 9\cdot 10^{17}$\bsep{1pt}\\
Held, He & 4030387200
\end{tabular}
\end{table}

We comment brief\/ly on some curious properties of the Monster group~$\mathbb M$. The Monster group has many singular features
\cite{Gan,Ron, Fre}. For example, (i) the prime decomposition of its
order (\ref{41}) contains 15 of the f\/irst 20 prime numbers; the
f\/irst one omitted is 37, and with the other four (43, 53, 61 and 67)
these are precisely (Oog) the f\/ive special primes for some {\it
modular functions} (see below) to represent the 2-sphere (and not
higher genus surfaces~\cite{Ron}).

(ii) The group $\mathbb M$ has 194 classes (of conjugate elements),
although the really independent ones are only 163, a remarkable
number (Gardner: Ramanujan), because $\exp(\pi\sqrt{163} )$ is very
nearly an integer~\cite[p.~227]{Ron}:
\begin{gather*}
\exp\big(\pi\sqrt{163} \big)=262537412640768743.99999999999925\dots.
\end{gather*}

Thus $\mathbb M$ has also 194 irreducible complex representations,
the three smallest having dimensions~1, $196 883 = 47\cdot 59\cdot
71$ and $21 296 876 = 2^2\cdot 31\cdot 41\cdot 59\cdot 71$. Notice
the smallest faithful {\it irrep} has dimensions 196883, close to
the number 196560, critical in the Leech lattice: this is one of the
hints relating the two (indeed the three) levels of sporadic groups.

This leads us \dots\ to the moonshine conjecture

(iii)  Another curious phenomenon, perhaps the more perplexing, was
discovered by McKay in 1980 and studied by others, named {\it
monstrous moonshine} (by Conway):

In a totally dif\/ferent domain of mathematics, namely the theory of
{\it elliptic modular functions}, there is a function $j(\tau )$ from the
upper complex plane $H$ (${\rm Im}\,\tau   > 0$) to the Riemann sphere $j( \tau )
: H \rightarrow\mathbb C^\sim$ ($= \mathbb C\cup \{ \infty\}$) whose
Laurent expansion reads (with $q= \exp(2\pi i\tau)$, $\tau\in H$,
$q(\tau +1) = q$)
\begin{gather}\label{44}
j(\tau ) = 1/q + 744 + 196 884q + 21 493 760q^2 + \cdots.
\end{gather}

Namely, after two terms, (to be expected) the coef\/f\/icients are given
by simple combination of the dimension of the {\it irreps} of the
Monster!; namely
\begin{gather*}
 196 884 = 1 + 196 883,\qquad  21 493 760 = 1+196 883 + 21 296
 876.
\end{gather*}

This coincidence was one of the most strange ever found in
mathematics! Understanding this has been a great break-through
(Frenkel, Borcherds), but we cannot explain it in this review. The
essential point is that there is a graded algebra, with graded
dimensions related to the dimensions of the {\it irreps} of the
Monster. The construction hinges on the theory of vertex operators
(appearing in string theory) and on a double generalization of the Lie
algebras (beyond the known Kac--Moody algebras), due to Borcherds
(1986).

The mystery has deepened: for other groups (e.g.~$E_8$) one has seen
also (V.~Kac) similar graded algebras; for a recent work on the ``moonshine'' for others finite groups, see~\cite{Harada}. Borcherds won the Fields
medal in 1998 for his work on $\mathbb M$.

\subsection{The Pariah groups}

The three levels of sporadic (f\/inite simple) groups have  certain
relations (the $5+7+8=20$ groups are called ``The happy family'' by R.~Griess \cite{Grie}), and in fact all of them can be considered as
subquotients of the Monster. But already the Janko group ${\rm J}_1$ is
not in these series; later analysis carried out in the period
1965--1975 ended up with 6 simple sporadic ``Pariah'' groups (name due
also to Griess), with no relation whatsoever with known groups
(except possibly two), thus completing the list of f\/inite simple
groups. We shall say nothing about them, but to give name and
orders
\begin{table}[h!]
\centering
\begin{tabular}{lr@{\,}l}
Rudvalis, Ru   &        $145926144000$ & $= 2^{14}\cdot
3^3\cdot 5^3\cdot 7\cdot 13\cdot 29 \approx  1.46\times 10^{11}$\bsep{1pt}\\
O'Nan, ON   & 460815505920 &\bsep{1pt}\\
Lyons,  Ly   & 51765179004000000 &\bsep{1pt}\\
Janko$_4$,  ${\rm J}_4$  &             86775571046077563880& \bsep{1pt}\\
Janko$_3$,  ${\rm J}_3$ &                   50232960 &\bsep{1pt}\\
Janko$_1$,  ${\rm J}_1$     &         175560 &
\end{tabular}
\end{table}

(The Janko group ${\rm J}_2$ is isomorphic to the Hall--Janko group HJ in
the second level.) The groups Ly and ${\rm J}_4$ have mappings into McL
(2nd level) and ${\rm M}_{24}$ (1$^{\rm st}$) respectively; the other four
are totally enigmatic at the moment (Jan.~2011); but there is also
a map of ON into ${\rm J}_1$, not yet understood.

\begin{figure}[h!]
\centerline{\includegraphics[width=159mm]{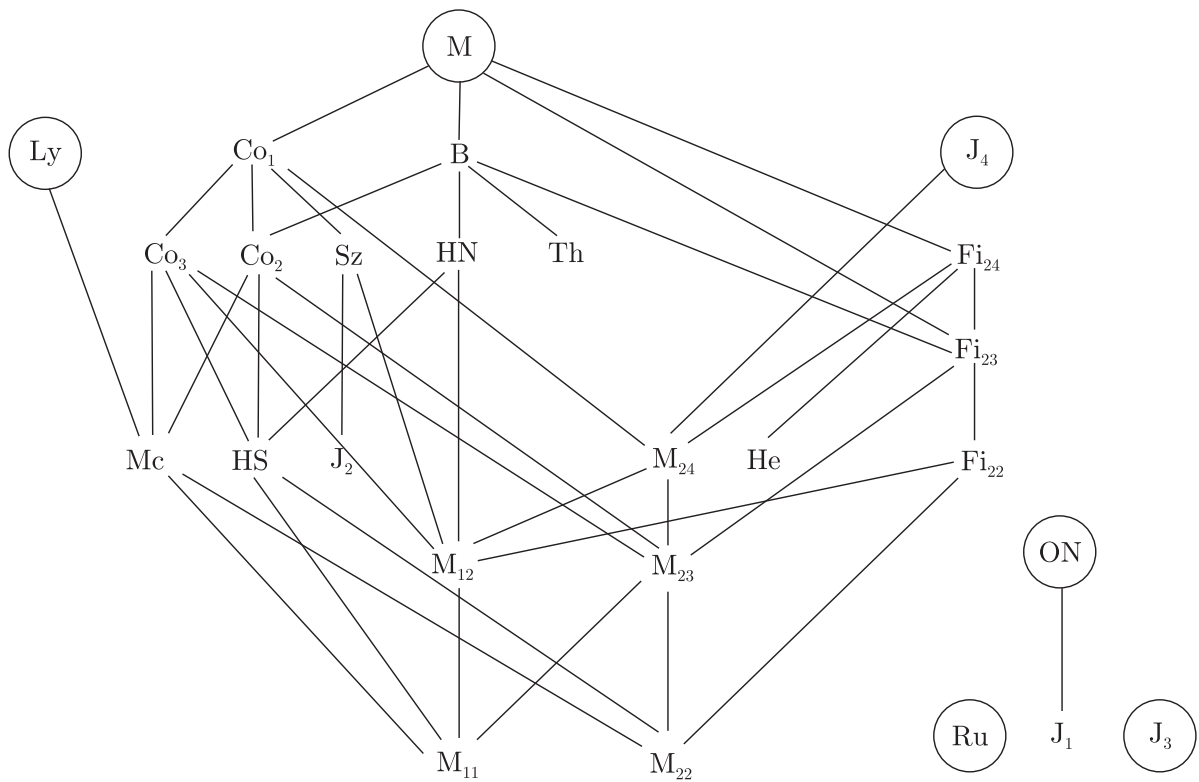}}
\caption{Genetic relation between sporadic f\/inite simple groups~\cite{Atl}.}\label{Fig1}

\end{figure}

So we end up by presenting the complete list of f\/inite simple
groups, giving only the group name (see Table~\ref{table1}).
On Fig.~\ref{Fig1} we
reproduce the page in the Atlas \cite{Atl} showing the genetic
relation between these groups.

\begin{table}[h!]
\centering
\caption{Finite simple groups {\it in families}.}\label{table1}
\vspace{1mm}

\begin{tabular}{llll}
\hline
1) & Abelian:       &&   $Z_p$,   $p$ any prime, order $p$\tsep{1pt}\\
2) & Alternative    &&           ${\rm Alt}_n$,  $n > 4$, order $n!/2$\\
3) & Lie-type (16) && \\
& & (1--4) Biparametric    & ${\rm PSL}_n(q)$, ${\rm PO}$, ${\rm PU}$,  ${\rm PSp}_n(q)$\\
& & (5--9) Uniparametric, Lie-type   &     $G_2(q)$, $F_4(q)$,  $E_{6,7,8}(q)$\\
&& (10--13) (Sternberg, autos) &    $^2A_n(q)$, $^2D_{n\ne 4}(q)$, $^2E_6(q)$, $^3D_4(q)$\\
&& (14--16)   (Suzuki, Rae, $\Longleftrightarrow$) &     $^2G_2(q)$,  $^2B_2(q)$, $^2F_4(q)$\\
4) & Sporadic groups & First level (5)  & ${\rm M}_{11, 12, 22, 23, 24}$ \\
&  & Second   (7)    &           ${\rm Co}_{1,2,3}$ -- HS, McL, HJ, Suz\\
 &&  Third   (8)   &     $\mathbb M$, ${\rm B}$, ${\rm Fi}_{22, 23, 24}$, Th, HN, He\\
 5) & Pariahs    (6)  &&                ${\rm J}_1$, ${\rm J}_3$, ${\rm J}_4$, Rud, ON, Ly\bsep{1pt}\\
 \hline
\end{tabular}
\end{table}

\section{Some applications of f\/inite simple groups in physics}

{\bf 1.~Elementary applications.} {\it Bose/Fermi statistics;
crystal groups.} Consider the Hamiltonian~$H$ for a system of $N$
electrons bound in a $Z$-protons nucleus
\begin{gather*}
 H = \sum(   p^2_i - Ze^2/r_i) + \sum_{i\ne j}
  e^2/|{\bf r}_i - {\bf r}_j| + \text{spin forces} + \text{relativistic corrections}+
  \cdots.
\end{gather*}

This Hamiltonian deals with {\it identical} electrons, so there is
an obvious symmetry $i \Longleftrightarrow j$ ($i, j: 1, \dots, N$).
In quantum mechanics symmetries are implemented usually by unitary
operators on the Hilbert space of states. So here we must have a
representation of the symmetric group~${\rm S}_N$. As the Jordan--H\"older
chain for ${\rm S}_N$ is  $\{ Z_2, {\rm Alt}_N\}$, the group has precisely two
1-dim {\it irreps}, the identical ${\rm Id} = D_0$ and the so-called
alternant $D_0^-$. The fundamental {\it spin-statistics
theorem} (Pauli, 1940) says that

``The symmetric group of a system of $N$ identical particles is
represented through the 1-dim {\it irreps}; indeed, integer spin
systems, called Bosons, are in the identical $D_0$, whereas
half-integer spin particles, called Fermions, chose the alternant
$D_0^-$ representations''.

In the second alternative $D_0^-$, we have the {\it exclusion
principle} as a consequence (Pauli, 1925), because electrons carry
spin $1/2$:

``There can be no more than one electron per orbital; or: two
electrons cannot have the same (four) quantum numbers''.

The importance of this is hard to overemphasize: the whole of
chemistry, up to ourselves, depends on the exclusion principle,
which classif\/ies electrons in orbitals, explains the chemical valence,
etc. It is amusing to speculate as on a~world, dif\/ferent from ours, in
which the isometry group of space will not be doubly connected: in
our case the rotation group ${\rm SO}(3)$ is not simply connected, so its
projective representations include those of the universal covering
group, ${\rm SU}(2)$: for half integer spin we have faithful {\it irreps}
of ${\rm SU}(2)$, as in the case of electrons. The exclusion principle
depends essentially on this:

``{\it The ultimate reason for the existence of the valence,
chemistry and ourselves is that the isometry group of our space is
not simply connected, allowing genuine projective $($half-integer
spin$)$ representations of the rotation group ${\rm SO}(3)$, which, in turn,
carry the alternant {\rm irrep} of the symmetric group, giving rise to
the exclusion principle}''.

As another application of f\/inite groups in physics we just mention
{\it passim} the classif\/ication of crystallographic groups, carried
out f\/irst by Bravais (ca.~1850), which uses the groups of elementary
polytopes  (cubic lattices etc.). For instance, the icosahedron
$Y_3$ has as rotation isometry the group ${\rm Alt}_5$; acting on the 12
vertices, the diagram explaining the situation is
\begin{gather*}
\begin{array}{@{}ccccc}
  Z_5 & \rightarrow & {\rm Alt}_5 & \rightarrow & Y_3(V)  \\
  \downarrow & \; & \downarrow & \; & \downarrow \\
 {\rm SO}(2) & \rightarrow & {\rm SO}(3) & \rightarrow & S^2
\end{array}
\end{gather*}
    Some viruses crystallize in icosahedra.

 {\bf 2.~Sphere packings, codes, etc.} Finite groups appear often
 in the sphere packing and kissing and covering problems; \cite{Con-2}
is the standard source. Related is the question of optimal
transmission of coded messages; the Leech lattice was discovered in
this context (1967). Sporadic groups in the f\/irst two levels are
very much related to these problems (e.g.\ the binary Golay code,
whose automorphism  group is again ${\rm M}_{24}$, see also \cite{Grie}),
but we do not elaborate on this engineering problem.

{\bf 3.~The Monster groups: string theory, vertex operators
and black holes.} For a~recent review of f\/inite groups in particle
physics, see~\cite{Ishi}. For relations between sporadic groups and
String theory see~\cite{Bor}.

The relation between the Monster and string theory (and/or
superstrings) will be  brief\/ly referred to here \cite{Fre,Gan,Ron}. An
expansion like (\ref{44}) was interpreted, from the Monster
($\mathbb M$) point of view, as a {\it graded} algebra, with a
(reducible) representation of $\mathbb M$ in each level. Analogous
things happen with Lie algebras $L$, there is an extension ({\it
affine} Kac--Moody algebras, 1964)~$L^{\widehat\,}$, whose natural
$\infty$-dim representation supports a graded representation of $L$~\cite{Kac}. Even for some Lie groups (like $E_8$) a ``moonshine
phenomenon'' appears~\cite{Gan,Harada}, as another modular function plays a~r\^ole. These generalized Lie algebras are very apt to describe
String Theory as a bidimensional conformal f\/ield theory embedded in
(25,1) dimensions, as in the primitive bosonic string theory; the
numbers $25-1=24 = 8\cdot 3$  always playing a role, see also~\cite{Bor}.

{\bf $\mathbb M$ and the BTZ 3-d black hole.} In 1993
Ba\~nados, Teitelboim and Zanelli (BTZ; from Chile) found a special
black hole appearing in 3-dim General Relativity {\it with a
negative cosmological constant} (as in 3-dim the Ricci tensor equals
the Riemann tensor, pure 3-dim gravitation as such is
uninteresting). In 2007 E.~Witten \cite{Witt} suggested a possible
relation between the Monster group $\mathbb M$ and the BTZ black
hole: the entropy of that hole might be related to the order of some
variable in the Monster; indeed, for the f\/irst faithful {\it irrep},
we have $d_2= 196 883$, which is related to number of states as
\begin{gather*}
\log (d_1 + d_2) = \log (196 884) = 12.190,
\end{gather*}
tantalizingly close to $4\pi=12.566$.

{\bf 4.  $\boldsymbol{{\rm M}_{24}}$ and the K3 surface.} Some recent work has
been triggered by \cite{Egu}, who discovered that the elliptic genus
of the K3 surface has a natural decomposition in terms of the
dimensions of irreducible representations of the largest Mathieu
group ${\rm M}_{24}$; about the K3 surface: this four-dim real manifold
is unique up to homeomophisms as carrying a ${\rm SU}(2)$ holonomy group,
thus being a Calabi--Yau 2-fold ${\rm CY}_2$. Now ${\rm CY}_3$ are much used in
compactif\/ication in string theories to pass from ten dimensions to
our realistic four, and it turns out that K3 plays a role there
also, see~\cite{Cheng}.

{\bf 5.  Overview.} We come to the end of our mini-review of
f\/inite simple groups; they appear in the abstract domains of superstrings
and black hole physics, as well as in some applied
physics regimes like transmission codes. One expects that these
surprising applications of an old domain of pure mathematics will be
increasing in the future \dots

\subsection*{Acknowledgements}

Work supported by grant A/9335/07 of the PCI-AECI and grant
2007-E24/2 of DGIID-DGA. The author thanks Professors A.~Andrianov
(Barcelona) and L.M.~Nieto (Valladolid) for the opportunity to
present the material as a Seminar in the Conference on
Supersymmetric Quantum Mechanics.

\pdfbookmark[1]{References}{ref}
\LastPageEnding


\begin{thebibliography}{99}

\footnotesize\itemsep=0pt


\bibitem{Bon}
Bonolis L.,
From the rise of the group concept to the stormy onset of group theory in the new quantum mechanics. A saga of the invariant characterization of physical objects, events and theories,
\href{http://dx.doi.org/10.1393/ncr/i2004-10006-4}{{\it Rivista del Nuovo Cimento}} {\bf 027}  (2004),   1--110.

\bibitem{Wey}
Weyl H.,
Theory of groups and quantum mechanics, Dover, New York, 1928.


\bibitem{Atl}
Conway J.H., Curtis R.T.,  Norton S.P., Parker  R.A.,  Wilson R.A.,
Atlas of f\/inite simple groups. Maximal subgroups and ordinary characters for simple groups,  Oxford University Press, Eynsham, 1985.


\bibitem{Thom}
Thomas A.D.,  Wood G.V.,
Group tables,  {\it Shiva Mathematics Series}, Vol.~2, Shiva Publishing Ltd., Cambridge, Mass., 1980.


\bibitem{Ram}
Ramond P.,
Group theory. A physicist's survey, Cambridge University Press, Cambridge, 2010.

\bibitem{Hal}
Hall M. Jr.,
The theory of groups, The Macmillan Co., New York,  1959.

\bibitem{Wig}
Wigner E.P.,
 Group theory and its application to the quantum mechanics of atomic spectra, {\it  Pure and Applied Physics},
  Vol.~5, Academic Press, New York~-- London, 1959.

  \bibitem{Rob}
Robinson D.J.S.,
 A course in the theory of groups, 2nd ed.,
 {\it Graduate Texts in Mathematics}, Vol.~80, Springer-Verlag, New York, 1996.

\bibitem{Bogo}
Bogopolski O.,
Introduction to group theory,  European Mathematical Society (EMS), Z\"urich, 2008.

\bibitem{Cox-1}
Coxeter H.S.M.,
Regular polytopes,  Dover Publications, Inc., New York, 1973.

\bibitem{Con-1}
Conway J.H., Smith D.A.,
On quaternions and octonions: their geometry, arithmetic, and symmetry,  A~K~Peters, Ltd., Natick, MA, 2003.



\bibitem{Mil}
Besche H.U., Eick B., O'Brien E.A.,
A millennium project: constructing small groups,
\href{http://dx.doi.org/10.1142/S0218196702001115}{{\it Inter\-nat. J. Algebra Comput.}} {\bf 12} (2002), 623--644.

\bibitem{Cox-2}
 Coxeter H.S.M., Moser W.O.J.,
Generators and relations for discrete groups, Springer-Verlag, New York~-- Heidelberg, 1972.

\bibitem{Car}
Carter R.W.,
Simple groups of Lie type, {\it Pure and Applied Mathematics}, Vol.~28, John Wiley \& Sons, London~-- New York~-- Sydney, 1972.

\bibitem{Art}
Artin E.,
Geometric algebra, Interscience Publishers, Inc., New York~-- London, 1957.

\bibitem{Dieu}
Dieudonn\'e J.,
La g\'eom\'etrie des groupes classiques,
Springer-Verlag, Berlin~-- G\"ottingen~-- Heidelberg, 1955.

\bibitem{Boy-1}
Boya L.J., Campoamor-Stursberg  R.,
Composition algebras and the two faces   of $G_2$,
\href{http://dx.doi.org/10.1142/S0219887810004348}{{\it Int. J. Geom. Methods Mod. Phys.}} {\bf 7} (2010), 367--378,
\href{http://arxiv.org/abs/0911.3387}{arXiv:0911.3387}.

\bibitem{Con-2}
Conway J.H., Sloane N.J.A.,
Sphere packings, lattices and groups,
{\it Grundlehren der Mathematischen Wissenschaften}, Vol.~290, Springer-Verlag, New York, 1988.

\bibitem{Grie}
Griess R.L. Jr.,
Twelve sporadic groups, {\it Springer Monographs in Mathematics}, Springer-Verlag, Berlin, 1998.

\bibitem{Why}
Biggs N.L., White A.T.,
 Permutation groups and combinatorial structures, {\it London Mathematical Society Lecture Note Series}, Vol.~33,
 Cambridge University Press, Cambridge~-- New York, 1979.

\bibitem{Boy-2}
Boya L.J.,
 New derivation of Mathieu groups, to appear.

\bibitem{Green}
Greenberg P.,
Mathieu groups, Courant Institute of Mathematical Sciences, New York University, New York, 1973.

\bibitem{Cox-3} Coxeter H.S.M.,
The beauty of geometry. Twelve essays,  Dover Publications, Inc., Mineola, NY, 1999.

\bibitem{Egu}
Eguchi T., Ooguri H., Tachikawa Y.,
Notes on the K3 Surface and the Mathieu group ${\rm M}_{24}$,
\href{http://arxiv.org/abs/1004.0956}{arXiv:1004.0956}.



\bibitem{Gan}
Ganon T.,
Monstrous moonshine: the f\/irst twenty-f\/ive years,
\href{http://dx.doi.org/10.1112/S0024609305018217}{{\it Bull. London Math. Soc.}} {\bf 38} (2006), 1--33.

\bibitem{Ron}
Ronan M.,
Symmetry and the Monster. One of the greatest quests of mathematics, Oxford University Press, Oxford, 2006.


\bibitem{Fre}
Frenkel I.,  Lepowsky J.,  Meurman A.,
Vertex operator algebras and the Monster, {\it Pure and Applied Mathematics}, Vol.~134, Academic Press, Inc., Boston, MA, 1988.

\bibitem{Harada}
Harada K.,
``Moonshine'' of f\/inite groups, {\it EMS Series of Lectures in Mathematics}, European Mathematical Society (EMS), Z\"urich, 2010.

\bibitem{Ishi}
Ishimori H.,  Kobayashi T.,  Ohki H.,  Okada H.,   Shimizu Y.,  Tanimoto M.,
Non-Abelian discrete symmetries in particle physics,
\href{http://dx.doi.org/10.1143/PTPS.183.1}{{\it Prog. Theor. Phys.}}  (2010),  suppl.~183, 1--163,
\href{http://arxiv.org/abs/1003.3552}{arXiv:1003.3552}.

\bibitem{Bor}
Borcherds R.E.,
Sporadic groups and string theory, in First European Congress of Mathematics, Vol.~I (Paris, 1992), {\it Progr. Math.}, Vol.~119, Birkh\"auser, Basel, 1994, 411--421.

\bibitem{Bor1}
Borcherds R.E.,
Book review: ``Moonshine beyond the Monster. The bridge connecting algebra, modular forms and physics'' by T.~Gannon,
\href{http://dx.doi.org/10.1090/S0273-0979-08-01209-3}{{\it Bull. Amer. Math. Soc.}} {\bf 45} (2008), 675--679.

\bibitem{Kac}
Kac V.G.,
 Inf\/inite-dimensional Lie algebras, 3rd ed., Cambridge University Press, Cambridge, 1990.

\bibitem{Witt}
Witten E.,
Three-dimensional gravity revisited,
\href{http://arxiv.org/abs/0706.3359}{arXiv:0706.3359}.

\bibitem{Cheng}
Cheng M.C.N., K3 surface, ${\mathcal N} = 4$ dyons, and the Mathieu group ${\rm M}_{24}$,
\href{http://arxiv.org/abs/1005.5415}{arXiv:1005.5415}.




\end{thebibliography}
\end{document}